\documentclass[11pt]{article}
\usepackage{rotating}

\usepackage{lscape}

\usepackage{graphicx}
\begin{document}
\title{\bf Galactic-Field Metal-Rich RR~Lyrae Variables: Features of Kinematics
and Abundances of Selected Chemical Elements }

\author{{M.\,L.~Gozha, V.\,A.~Marsakov,  V.\,V.~Koval'}\\
{Southern Federal University, Rostov-on-Don, Russia}\\
{e-mail:  gozha\_marina@mail.ru, marsakov@sfedu.ru, litlevera@rambler.ru}}
\date{accepted \ 2024, Astrophysical Bulletin, Vol. 79, No. 3, pp. ??-???}

\maketitle

\begin {abstract}
We present an analysis of the relative abundance features of a number 
of chemical elements in the atmospheres of metal-rich ($\rm{[Fe/H]} > -1.0$)
Galactic-field RR~Lyrae variable stars and the kinematic characteristics
of these stars. We have previously shown that the relative abundances of
some $\alpha$-elements: magnesium, silicon, calcium, and to a greater 
extent of titanium, as well as yttrium and scandium in such 
stars are lower than in most of other types of stars, bearing similar 
metallicity. It is found here that some of these metal-rich RR~Lyrae 
stars also have very low relative abundances of sodium, aluminum and 
nickel. The orbital parameters of all the metal-rich RR~Lyrae variables 
studied in this paper are typical of the Galactic thin or thick
disk objects, however, unusual chemical composition 
let us to suggests a possible extragalactic origin for some of them.

\end{abstract}


\maketitle

\section{INTRODUCTION}

RR~Lyrae variable stars are radially pulsating variables of 
spectral classes A -- F at the helium burning
core stage of evolution, residing on the horizontal
branch in the Hertzsprung-Russell diagram. Stars at
such an advanced stage of evolution are traditionally
considered to be an exclusively old and metal-poor
stellar population. However, among the RR~Lyrae
stars of the Galactic field there proved to be a small
number of stars with an approximately solar metallicity. Such a 
high metallicity contradicts the modern
model of stellar evolution, according to which only the
low-metallicity low-mass stars get to the horizontal
branch in the instability strip, moreover, they evolve
to it for more than 10 billion years, i.\,e. longer than
the lifetime of the Galactic thin disk subsystem, which
is characterized by this high metallicities. Note that
RR~Lyrae variable stars are practically not observed
in metal-rich globular clusters, since their horizontal
branches lie entirely in the region of temperatures
lower than those in the instability strip.

In this paper, we carry out new research of
RR~Lyrae stars of the Galactic field, initiated by
Marsakov et\,al.~(2018) and continued by 
Marsakov et\,al.~(2019) and Gozha et\,al.~(2020). In these
studies, we used the authors' catalog containing the
data on the positions, velocities, and metallicities for
415 field RR~Lyrae stars. For a hundred of these stars,
information on the relative abundances of several
chemical elements obtained from high-resolution
spectra was found in the literature. It was shown that
most of the metal-rich ($\rm{[Fe/H]} > -1.0$) RR~Lyrae stars
with lowered abundances of magnesium, silicon,
calcium, titanium, scandium, and yttrium have the
kinematics characteristic of the youngest subsystem,
namely, the Galactic thin disk. Interestingly, the
[Mg/Fe], [Si/Fe], and [Ca/Fe] ratios in the metal-rich 
RR~Lyrae stars are lower than the average ratios
for field stars, while the [Sc/Fe], [Ti/Fe], and [Y/Fe]
values are generally even lower than in all thin-disk
stars. At the same time, no traces of such elements
as europium, zirconium, and lanthanum, which are
present in all thin-disk stars, were detected in these
stars. According to our stratification, some metal-rich 
RR~Lyrae stars belong to the thick disk according
to their kinematics.

The analysis of the RR~Lyrae star kinematics in
our earlier papers was carried out using the values
of the spatial velocity components calculated based
on the data from the catalog by Dambis~et\,al.~(2013), 
which used ground-based measurements of
the proper motions of stars. In this paper, we test
our earlier conclusions using the new, more accurate
astrometric and photometric satellite data.

Hasselquist et\,al.~(2021) studied the chemical
composition of stars in nearby dwarf galaxies and
found a deficit of some chemical elements. Among
them were the elements that we have not yet analyzed
in the RR~Lyrae stars. In this regard, we considered it 
necessary to continue studying the chemical
and kinematic properties of RR~Lyrae variables of the
galactic field, and above all the metal-rich stars.


\section {INITIAL DATA}

The spectroscopic measurements of [Fe/H] for
100 Galactic-field RR~Lyrae variables, as well as the
[Al/Fe] ratios for 44 of them were taken from the authors' 
compiled catalog (Marsakov et\,al.~(2018)). It
consolidates the data on the abundances of chemical
elements from 25~publications from 1995 to 2017.
The relative abundances of chemical elements were
calculated as weighted averages in cases of two or
more measurements by different authors and brught 
to an unified solar abundance. The standard deviation of the 
calculated means: $\sigma_{\rm{[Fe/H]}} = 0.12$, 
$\sigma_{\rm{[Al/Fe]}} = 0.17$. 
Detailed information about the catalog is given in the
paper by Marsakov et\,al.~(2018)\footnote{The catalog can be found at: 
http://vizier.u-strasbg.fr/viz-bin/VizieR?-source=J/AZh/95/54}.

Hasselquist et\,al.~(2021) found a deficit in the
relative abundances of aluminum and nickel in the
stars of the dwarf satellite galaxies. The data on
the relative abundances of nickel [Ni/Fe] obtained
spectroscopically was found in the literature for 55
RR~Lyrae stars from our catalog. The data were collected from 
9~papers from 1995 to 2015. To determine
the [Ni/Fe] values, we performed procedures similar
to those used to obtain the relative abundances of
chemical elements when compiling the catalog of
Marsakov et\,al.~(2018). In particular, we brught the
authors' [Ni/Fe] values for each RR~Lyrae star to an unified  
solar abundance from Asplund et\,al. (2009)
and calculated their weighted averages. Since the
[Ni/Fe] abundances obtained by different authors for
one and the same star are scarce, the standard deviation of the
weighted average was not calculated. The average
error calculated taking into account the uncertainties
of the [Ni/Fe] values adopted from the literature was
0.12~dex. However, it should be noted that this value
is actually an underestimate, since only slightly more
than a half of the [Ni/Fe] values in the literature were
given with uncertainties.

In addition, this paper analyzes the relative abundances of 
sodium, the synthesis of which is largely
analogous to the process of aluminum formation. Homogeneous 
data on the sodium abundances [Na/Fe]
in the atmospheres of 18 RR~Lyrae stars, determined
from high-resolution spectra are adopted from the
recent paper of Takeda (2022). The error declared by
the author does not exceed 0.2~dex.

The catalogue of Prudil et\,al.~(2020), based on the
satellite measurements of distances and velocities,
served as a source of orbital parameters for 314
RR~Lyrae variables. Of these, the authors selected
22 RR~Lyrae stars, which, in their opinion, belong
to the disk subsystems of the Galaxy, based on two
arbitrarily chosen strict orbital criteria, namely, small
orbital eccentricities ($e < 0.2$) and rather small maximum height 
of the stellar orbit points from the Galactic
plane ($Z_{max}<0.9$~kpc). It was shown that the set
of selected stars is characterized by small velocity
dispersions ($\sigma_{V_{tot}} = 37$~km\,s$^{-1}$ 
and $\sigma_{V_{z}} = 16$~km\,s$^{-1}$
for the total velocity and its vertical component,
respectively), high metallicity (the average value of
$\rm{[Fe/H]_{aver}} = -0.60$) and around solar abundances of
$\alpha$-elements (for nine stars with known calcium abundances, the 
average value of $\rm{[Ca/Fe]_{aver}} = 0.02$),
which, according to the authors of the paper, indicates
that these stars may belong to the thin disk.

We used several samples of field stars with known
metallicities and the relative abundances of aluminum,
sodium, and nickel, obtained from the high-resolution
spectra as comparison objects. From the
APOGEE DR\,16 catalogue (J\"onsson et\,al.,~2020), 
including over four hundred thousand stars of the
Galaxy, we selected 13 828 stars with the chemical
composition data available and with the atmospheric
parameters identical to the characteristics of the
atmospheres of RR~Lyrae stars. The data on the
metallicity and relative abundances of aluminum,
sodium, and nickel in 1918 red giants were taken
from the catalogue of Hawkins et\,al.~(2016). The
sources of data on [Fe/H], [Al/Fe], [Na/Fe], and [Ni/Fe] in the
atmospheres of variable stars were: the paper by Luck(2018)
for 435~Cepheids, and the paper by Kovtyukh et\,al.~(2018) for 23 
type II Cepheids (W Virgo and
BL~Hercules types).

\section {ORBITAL PARAMETERS}

Marsakov et\,al.~(2018) showed that a
significant fraction of metal-rich RR~Lyrae stars
move in the Galaxy similarly to thin-disk objects. To
stratify the RR~Lyrae stars by Galactic subsystems,
we used the technique proposed by Bensby et\,al.~(2003). 
It calculates the probability of field stars to
belong to the thin, thick disk or the halo subsystems
based on the components of their spatial velocities
relative to the local centroid and the dispersion of
these components in each subsystem.

To confirm that metal-rich RR~Lyrae variables belong 
to the Galactic disk, let us discuss some of
their orbital characteristics obtained from the satellite
data. The catalog of Prudil et\,al.~(2020) presents the
kinematic and orbital parameters for 314 RR~Lyrae
variables. The [Fe/H] values for 68 of them are
contained in the catalog of Marsakov et\,al.~(2018).

Figure~1a,\,b shows the distribution of the field RR~Lyrae
variables from the catalogue of Prudil et\,al.~(2020) based on the 
orbital parameters. Figure~1a demonstrates 
the ``eccentricity $e$ --- maximum
distance of the star's orbit points from the Galactic
plane $Z_{max}$'' diagram. For convenience, we limited
the vertical axis to 30~kpc; two stars out of the 314
RR~Lyrae stars in the Prudil et\,al.~(2020) were
left outside the diagram area. Figure 1b shows the
``apogalactic distance $R_{apo}$ --- maximum distance  of
the star's orbit points from the Galactic plane $Z_{max}$''
diagram for stars in the Prudil et\,al.~(2020) catalog.
Three stars with the largest apogalactic distances
were left outside the boundaries of Figure~1b.

Note that out of 22 RR~Lyrae stars labeled in
Prudil et\,al.~(2020) as the disk subsystem stars, in our
earlier paper (Marsakov et\,al.~(2018)) 21 variables were
assigned to the thin disk population using the spatial-kinematic 
probability criterion, while FH~Vul was
referred to the thick disk. For this variable, $Z_{max}$ and $e$
are close to the upper limit values taken up in Prudil et\,al.~(2020) 
for disk stars. Hence, we observe a match
of the stratification results for RR~Lyrae variables in
the two studies. At that, in the paper 
of Marsakov et\,al.~(2018), 
the criteria were the components of
spatial velocities and their dispersions, 
and Prudil et\,al.~(2020) 
used the orbital parameters obtained from
the satellite data.

We can see on the diagrams of Fig.~1a, b that
all the stars of disk subsystems according to Prudil et\,al.~(2020) 
with known spectroscopic metallicity
measurements (there are nine such RR~Lyrae stars)
turned out to be metal-rich.

We decided to verify the results of stratification of
metal-rich RR~Lyrae stars into Galactic subsystems,
obtained earlier in Marsakov et\,al.~(2018) using
the up-to-date satellite data. For this purpose, we
used the astrometric parameters of the Gaia~DR3
catalog. The radial velocity data for the vast majority
of studied stars are also present in this catalog. The
radial velocity for one RR~Lyrae star was found in the
Gaia~DR2 catalog, and for three more stars -- in the
SIMBAD database. The spatial velocity components
were determined taking into account the position and
velocity of the Sun used by Prudil et\,al.~(2020). In this
case, to identify the belonging of the RR~Lyrae stars to
different subsystems, we also applied the probabilistic
method from the work of Bensby et\,al.~(2003). Out of
25 metal-rich RR~Lyrae stars, only TV~Lib, assigned
by Marsakov et\,al.~(2018) to the thin disk population,
turned out to be a thick disk star according to the
new calculations. The conclusion of Marsakov et\,al.~(2018) 
on the membership of the remaining metal-rich 
RR~Lyrae stars in the galactic subsystems was
confirmed.

Based on the updated stratification by subsystems, 
we find that out of 17 RR~Lyrae variables with
$\rm{[Fe/H]} > -1$ of the thin disk, only 9~stars fell into
this subsystem according to the strict orbital criterion 
of Prudil et\,al.~(2020). Hence, the RR~Lyrae
stars SW~And, RS~Boo, DM~Cyg, XZ~Dra, TW~Her,
CN~Lyr, V~445~Oph, AR~Per and HH~Pup with the
spectroscopically determined metallicity of $\rm{[Fe/H]} >-1$ 
can be recognized as the thin galactic disk objects
based on their kinematic and orbital characteristics.
For five other RR~Lyrae stars that we have assigned
to the thin disk according to the probabilistic criteria,
only one of the two orbital parameters shows a slight
excess of the limiting values, adopted in Prudil et\,al.~(2020). 
Since the limiting values specified in the
latter study are not well defined mathematically, the
stars DX~Del, RR~Gem, KX~Lyr, AV~Peg and AN~Ser
can also be assigned to the thin disk. The orbital
parameters for the three metal-rich stars assigned to
the thin disk in this study have not been calculated.

Seven metal-rich RR~Lyrae stars, identified in
this paper as thick disk objects, have eccentricities
of $e = 0.19 - 0.45$ and move away from the Galactic
plane at the distances of up to $Z_{max} = 0.60 - 2.00$~kpc
(according to the catalog by Prudil et al., 2020). Such
orbital parameters are quite consistent with the characteristics 
of thick disk objects. Another one RR~Lyrae
star with $\rm{[Fe/H]} > -1$ that we have classified as a
halo star based on its kinematics, is missing in the
catalog of Prudil et\,al.~(2020).

\section {CHEMICAL COMPOSITION}

Various chemical elements are produced by stars of different 
masses, consequently they enter the interstelar medium on 
different time scales. The chemical
composition of stellar atmospheres hence reflects the
history of the chemical evolution of matter from which
they have formed. Therefore, each galaxy has a
unique history of metal enrichment depending on the
mass of the galaxy, the rate of star formation, and the
masses of supernovae exploding in it.

We are now aware of numerous events when stellar
and gaseous matter of dwarf galaxies was captured
by the tidal forces of our Galaxy at
different times of its evolution (see, e.\,g., Kruijssen
et\,al.,~2020; Naidu et\,al.,~2020). It hence seems
intriguing to identify stars in our Galaxy with chemical 
composition features that would differ from those,
typical of objects formed during the evolution of a
single protogalactic cloud. Since the star formation
rate in larger galaxies is higher than that in
dwarf galaxies (see, e.g., Matteucci and Greggio,
1986), the abundances of some chemical elements in
the stars born in massive galaxies may differ from the
chemical composition of stars, captured from dwarf
satellite galaxies.

RR~Lyrae stars are traditionally considered to be
an old population of the Galaxy; the vast majority
of these stars have low metallicity. However, some
RR~Lyrae stars have metallicities, uncharacteristically 
high for old stars ($\rm{[Fe/H]} > -1$). Early studies
showed that the relative abundances of some chemical elements 
in such RR~Lyrae stars differ from the abundances
of the same elements in stars with similar metallicity. 
In that way, previous studies of Marsakov et\,al.~(2018), 
Gozha~et\,al.~(2020, 2021) drew attention 
to the lack of magnesium, silicon, calcium,
and especially scandium, titanium, and yttrium in the
atmospheres of metal-rich RR~Lyrae stars, compared
to field stars of similar metallicity (the comparison
was also made with variable stars).

Hasselquist et al. (2021) studied the abundances
of several chemical elements, determined from high-resolution 
spectra in the atmospheres of red giants
in massive satellites of the Milky Way: in the Large
and Small Magellanic Clouds, the Sagittarius and
Fornax Dwarf Spheroidal Galaxies, and in the Gaia-Enceladus 
system, which is by now fully disrupted.
These authors compared the abundances of chemical
elements in the dwarf galaxies with the corresponding
abundances in the stars of the disk and bulge of
our Galaxy. The studied elements include the 
$\alpha$-elements, as well as [Al/Fe] and [Ni/Fe]. In the
[X/Fe] vs. metallicity diagrams (Fig.~5 in Hasselquist et al., 2021), 
the stars of each of the studied
dwarf galaxy occupy the regions that practically do
not overlap with the stars of the Milky Way. The
``[Fe/H] -- [Al/Fe]'' and ``[Fe/H] -- [Ni/Fe]'' diagrams
are particularly prominent: the relative abundances
of aluminum and nickel in the stars of dwarf satellite 
galaxies are lower than those in the stars of the
Galaxy at any metallicity. The authors explain this
fact by the differences in the histories of chemical
evolution in a particular galaxy.

Feuillet et\,al.~(2022) in their paper discuss the
possible presence in the galactic disk of old 
high-metallicity stars, captured by our Galaxy during the
mergers with satellite galaxies. The paper considers
a sample of giant stars, where the authors identify
stars with unusual chemical compositions and thin-disk 
kinematics. The authors classify such objects
as accreted. They select stars with anomalously low
aluminum abundances from the disk objects (i.\,e. with
metallicity $\rm{[Fe/H]} > -0.8$ and velocity around the
galactic center $V > 110$~km\,s$^{-1}$). This is reasoned
by the fact that the relative aluminum abundances
in the stars of dwarf galaxies, the satellites of the
Milky Way, are actually lower. From nine
metal-rich RR~Lyrae variables with known [Al/Fe] and
spatial velocities, Feuillet et\,al.~(2022) selected the
stars with thin-disk kinematics and equally low relative 
aluminum abundances ($\rm{[Al/Fe]} < -0.14$) as in
the accreted red giants of the disk. According to
the authors, five RR~Lyrae stars satisfy these criteria:
CN~Lyr, DM~Cyg, DX~Del, RS~Boo, V~445~Oph. The
authors propose to classify AA~Aql with low [Al/Fe]
but with the thick disk (or halo) kinematics and, 
possibly, SW~And, for which the [Al/Fe] value is slightly
higher than the arbitrary limiting value adopted in the
paper, as accreted stars.

We also noted earlier the possible extragalactic origin of
metal-rich field RR~Lyrae variables 
in (Marsakov et al., 2018, 2020) based on the under-estimated 
abundances of $\alpha$-elements in them.

\subsection {Aluminum, Sodium, and Nickel Abundances}

Let us consider the nature of metallicity 
dependences of the relative abundances of aluminum,
sodium, and nickel in the atmospheres of the field
RR~Lyrae variables from our catalog. We will be primarily 
interested in the abundances of these elements
in metal-rich RR~Lyrae stars.

Aluminum and sodium are light elements with an
odd number of protons. The main mechanism for
producing these elements is the hydrostatic burning of carbon 
for sodium, and carbon and neon for
aluminum in the cores of massive stars. Additional
synthesis is possible in the neutron capture reactions
occurring simultaneously with the CNO cycle: by
magnesium atoms in the cores of intermediate-mass
stars for aluminum, and by neon atoms in the cores
of intermediate-mass stars and during the hydrogen
burning in their layer sources for sodium. Sodium and
aluminum, produced in stars via similar processes,
show resembling trends in the ``[Fe/H] -- [Na/Fe]''
and ``[Fe/H] -- [Al/Fe]'' diagrams (see, for example,
Figs.~18 and~19 in Kobayashi et al., 2020). In
addition, models predict that the relative abundances
of aluminum and sodium decrease with increasing
metallicity at $\rm{[Fe/H]} > -1$, which is associated with
the onset of SNIa outbursts (Hawkins et al., 2016).
Since aluminum and sodium are formed in almost
identical processes, it seems worthwhile to analyze
not only the aluminum abundances but also those
of sodium in metal-rich RR~Lyrae stars.

Nickel is an iron-peak element. This element, like
iron, is produced and dissipated into the interstellar
medium in large quantities in type Ia supernovae
(see, e.g., Iwamoto et\,al.,~1999), and some nickel is
produced in type II supernovae (see, e.g., Kobayashi
et\,al.,~2006).

Hasselquist et al. (2021) showed that at any
metallicity, the [Al/Fe] and [Ni/Fe] ratios in the stars
from the studied dwarf galaxies are lower than the
abundances of these elements in the stars of the
Galaxy. Feuillet et\,al.~(2022) analyzed the aluminum
abundances in metal-rich RR~Lyrae stars of the field of
our Galaxy in order to identify the accreted stars. To
do this, they used the [Al/Fe] values from Liu et\,al.~(2013). 
In our catalog (Marsakov et\,al.,~2018), the
relative abundance of aluminum in RR~Lyrae stars
was determined using the information from a large
number of sources, the data were averaged and 
brought to an unified solar abundance. This is why
it will now be useful to compare the conclusions of
Feuillet et\,al.~(2022) with the results of analysis of the
Marsakov et\,al.~(2018) catalog data, and also to consider 
the nickel abundance vs. metallicity relationship
in the RR~Lyrae stars.

To study the behavior of the nickel abundance,
we determined the [Ni/Fe] ratios in the atmospheres
of 55 RR~Lyrae variables using the literature data.
Table 1 presents the collected and averaged values of
the relative nickel abundance, as well as [Fe/H] from
the Marsakov et\,al. catalog~(2018) and references to
data sources for [Ni/Fe].

Table 2 contains metallicities and relative abundances 
of the elements we study for 14 metal-rich
RR~Lyrae variables. The [Fe/H], [Al/Fe], [Ti/Fe], and
[Y/Fe] values are taken from Marsakov et\,al.~(2018),
[Na/Fe] -- from Takeda~(2022), [Sc/Fe] -- from 
Gozha et\,al.~(2020). The [Ni/Fe] values are found in this
paper.

\subsection {[Al/Fe], [Na/Fe], [Ni/Fe] vs. Metallicity
Relationship in RR~Lyrae Stars and Other Types of
Stars}

Figure 2 shows the ``[Fe/H] -- [Al/Fe]'' (panels
a, b, c), ``[Fe/H] -- [Na/Fe]'' (panels d, e, f) and
``[Fe/H]~-- [Ni/Fe]'' (panels g, h, i) diagrams for the
RR~Lyrae variables and comparison stars. All diagrams 
show the RR~Lyrae variables with known relative 
abundances of the considered chemical elements.
The following literature data were used: [Fe/H],
[Al/Fe] values are taken from the Marsakov et\,al.
catalog~(2018); [Na/Fe] data for 18 RR~Lyrae stars
are adopted from Takeda~(2022), their list contains
only a few metal-poor RR~Lyrae stars (all their
stars have $\rm{[Fe/H]} > -1.5$). The [Ni/Fe] ratios are
determined in the present paper.

Along with the analysis of the behavior of [Al/Fe],
[Na/Fe], and [Ni/Fe] relative to the metallicity 
variations for the full sample of RR~Lyrae stars, 
we compare the relative abundances of chemical elements
studied in the atmospheres of metal-rich RR~Lyrae
variables and Galactic field stars. Let us compare with the
abundances of aluminum, sodium, and nickel in the
atmospheres of red giants, classical Cepheids, and
type II Cepheids (W~Virgo and BL~Hercules). Comparison 
stars have been added to all the panels of Fig.~2
 with diagrams showing the relationship between
the relative abundances of aluminum, sodium, and
nickel vs. metallicity. The choice of comparison
objects was determined by the close luminosities and
temperatures of these stars and RR~Lyrae variables.
At that, the objects with which the RR~Lyrae stars
were compared are both stationary (red giants) and
variable (young classical Cepheids and old type II
Cepheids) stars.

Trend lines, depicting the variations in the relative
abundances of elements under study with increasing
metallicity were drawn at the diagrams of Fig.~2 for
the metal-rich RR~Lyrae variables (here we did not
take into account TV~Lib and KP~Cyg, which deviate 
from the general trend) and for the comparison
stars. In all panels of Fig.~2 (except Fig.~2g), linear
regressions were constructed using the least squares
method. In the ``[Fe/H] -- [Ni/Fe]'' diagram (Fig.~2g),
a second-order polynomial approximation was used
for the red giants.

Fig. 2a, b, c show the relationship between the relative 
abundance of aluminum and metallicity in the atmospheres 
of RR~Lyrae variables. The abundance of aluminum in the metal-poor
RR~Lyrae stars varies over a wide range, while the
metal-rich RR~Lyrae stars (except for TV~Lib
and KP~Cyg) show a small scatter of [Al/Fe] values
of 0.2~dex. The relative abundance of aluminum in
the atmospheres of the metal-rich RR~Lyrae stars
is below the solar value. In the RR~Lyrae stars
with $\rm{[Fe/H]} > -1$, [Al/Fe] decreases with increasing
[Fe/H]. In all the ``[Fe/H] -- [Al/Fe]'' diagrams, we
clearly see that the metal-rich RR~Lyrae stars lie
below the regions occupied by comparison stars of
the same metallicity (except for the already mentioned
TV~Lib and KP~Cyg).

The RR~Lyrae variables in the ``[Fe/H] -- [Na/Fe]''
diagrams (Fig.~2d,\,e,\,f) are predominantly metal-rich. 
Here, as in the case of aluminum, a tendency
for [Na/Fe] to decrease with increasing metallicity
is observed in the range of metal-rich RR~Lyrae stars.
Nevertheless, five RR~Lyrae stars with $\rm{[Fe/H]} > -1$,
including TV~Lib, have sodium abundances exceeding 
the solar value. For another star with
an aluminum excess, KP~Cyg, the relative sodium
abundance was not determined in Takeda (2022).
However, the sodium abundance in the atmosphere
of KP~Cyg is listed in the catalog of Marsakov et~al.~(2018), 
and it is also high ([Na/Fe] = 0.31).
Among the metal-rich RR~Lyrae stars, the lowest
sodium abundance ($\rm{[Na/Fe]} = -0.23$) is observed
in the atmosphere of CI~And (however, for this star
$\rm{[Ni/Fe]} = 0.06$ is close to the average value for
metal-rich RR~Lyrae stars, but the aluminum
abundance was not determined). The red giants
demonstrate a trend, similar to the RR~Lyrae stars,
namely, a decreasing relative sodium abundance with 
increasing metallicity (Fig.~2d), while the metal-rich 
RR~Lyrae stars prove to be almost in the middle of
the giant sequence. That is, there are no differences
in the relative sodium abundances between these
objects. This fact can be explained assuming that
accreted objects are abundant in the star sample of
the Hawkins et\,al. catalog~(2016). Unfortunately,
there is no information about their belonging to
galactic subsystems or about the spatial positions
and velocities of the stars in the paper of 
Hawkins et\,al.~(2016). Nevertheless, it can be seen that in
the ``[Fe/H] -- [Na/Fe]'' diagram (Fig.~2d), the most
metal-rich RR~Lyrae stars still lie slightly below the 
average line for red giants. Furthermore, in Fig.~2e, it is
clearly visible that the sodium abundances in metal-rich 
RR~Lyrae stars are significantly lower than in the
classical Cepheids. A similar situation is observed for
the sodium abundance when compared with Type II
Cepheids, although with a smaller difference (Fig.~2f).

In the of Fig.~2g,\,h,\,i with the
``[Fe/H] -- [Ni/Fe]'' diagrams, a significant scatter of
relative nickel abundances is observed (similar to
aluminum) at the metallicity $\rm{[Fe/H]} < -1$. However,
14 metal-rich RR~Lyrae stars with known [Ni/Fe]
ratios form a fairly compact group in the diagram
with a decreasing nickel abundance at increasing
metallicity. This trend to a greater extent applies
to metal-rich stars with $\rm{[Ni/Fe]} < 0$. Let us
pay attention to TV~Lib and KP~Cyg, whose relative
nickel abundances exceed those of the Sun (in the
diagrams, KP~Cyg is located far from the compact
group of metal-rich RR~Lyrae stars, while TV~Lib
has the highest [Ni/Fe] value for the RR~Lyrae stars
with $\rm{[Fe/H]} > -1$). The nickel abundances in metal-rich 
RR~Lyrae variables and comparison stars do not
show a significant difference, but here as well, about
a half of the RR~Lyrae stars with $\rm{[Fe/H]} > -1$ in the
``[Fe/H] -- [Ni/Fe]'' diagrams lie below the regression
lines for the comparison stars. Hawkins et\,al.~(2016)
noted that red giants are on the average located on
the ``[Fe/H] -- [Ni/Fe]'' diagram along the horizontal
line with a weak tendency for [Ni/Fe] to increase with
increasing metallicity at $\rm{[Fe/H]} > 0$ (see our Fig.~2g).
However, we do not observe this same tendency,
noted in the work of Hawkins et\,al.~(2016) for a large
sample of red giants, for our metal-rich RR~Lyrae
stars in the ``[Fe/H] -- [Ni/Fe]'' diagrams. In contrast,
the metal-rich RR~Lyrae variables (except for
KP~Cyg alone) clearly demonstrate a decrease in
[Ni/Fe] values with increasing metallicity. Note that
in the case of nickel (as well as for sodium), the metal-rich 
RR~Lyrae stars lie inside the red giant region.

Let us add a comparison of the behavior of
metal-rich RR~Lyrae stars relative to the stars
from more representative samples. Fig.~3 shows
the ``[Fe/H] -- [Al/Fe]'' (a) and ``[Fe/H] -- [Ni/Fe]''
(b) diagrams for RR~Lyrae variables from the catalog 
of Marsakov et\,al.~(2018) and stars from
the extensive APOGEE~DR16 catalog by J\"onsson et\,al.~(2020). 
We made a selection of data from
the APOGEE~DR16 catalog: from almost half a
million stars of the Galaxy, we selected 13 828 stars
with the chemical composition data, based on the
$T_{eff} = (5900-7700)$~K and $\rm{log} g \leq 4.2$ criteria, so
that these atmospheric parameters were close to the
characteristics of the atmospheres of RR~Lyrae
variables. The overwhelming majority of the stars that
we have selected for comparison turned out to have
the same metallicity as metal-rich RR~Lyrae
variables. Fig.~3a,\,b clearly show metal-rich
RR~Lyrae stars, except for TV~Lib and KP~Cyg, are
located below the regions of increased density of the
J\"onsson et\,al. catalog~(2020) stars.

We see from Figs.~2 and~3 (and also from Table~2)
that a significant fraction of metal-rich RR~Lyrae
variables have aluminum, sodium, and nickel abundances 
that do not exceed the solar values. The
RR~Lyrae stars with known abundances of all six
chemical elements from Table~2 that satisfy the
$\rm{[X/Fe]} \leq0.0$ criterion are SW~And, RS~Boo, DM~Cyg
and CN~Lyr.

Therefore, the relative abundances of aluminum,
sodium, and nickel in most metal-rich RR~Lyrae variables 
are significantly lower than the average [Al/Fe],
[Na/Fe], and [Ni/Fe] values in the comparison stars,
with the most noticeable difference observed for 
aluminum. However, one should be cautious about
the results of comparison of the metal-rich RR~Lyrae
stars with red giants for the abundances of sodium
and nickel. It is not possible to explain the low
values of [Al/Fe], [Na/Fe], and [Ni/Fe] in metal-rich
RR~Lyrae variables by the effects arising in the 
non-stationary atmosphere of these stars, since the 
comparison was made with both stationary and variable
field stars.

We therefore observe anomalously low relative
abundances of magnesium, silicon, calcium, 
scandium, titanium, yttrium (see our earlier works),
aluminum, sodium, and nickel in the atmospheres
of a significant number of metal-rich RR~Lyrae
variables compared to the field stars of the 
corresponding metallicity. And it is nessesary  
to try to find
an explanation for this.

\section {DISCUSSION OF THE PROPERTIES 
OF METAL-RICH RR LYRAE STARS}

Let us consider the properties of individual stars.
In the atmospheres of DM~Cyg and CN~Lyr, the
abundances of all nine chemical elements (magnesium, 
silicon, calcium, titanium, yttrium, scandium,
aluminum, sodium, and nickel) studied in this and
our previous papers are lower than in most of other
metal-rich RR~Lyrae stars and in the comparison 
stars of similar metallicity. These stars exhibit
the thin-disk kinematics (this paper, Prudil et al.,
2020). Only because of their low [Al/Fe] values,
Feuillet et al. (2022) considered them to have been
captured from the satellite galaxies. The RR~Lyrae
star V~445~Oph, identified by Feuillet et al. (2022)
to be accreted, with low abundances of five chemical
elements from Table 2 (except [Na/Fe]), has the thin-disk 
kinematics (this paper; Prudil et al., 2020). The
star RS~Boo, classified as disk star by its kinematics
(this paper; Prudil et al., 2020) and accreted by its
chemical composition (Feuillet et al., 2022), shows
low abundances of four chemical elements (except
for sodium and nickel abundances, close to solar
values). Eight more RR~Lyrae stars from Table 2
(except for KP~Cyg and TV~Lib) contain anomalously
low abundances of some of the nine chemical elements 
considered. Hence, twelve of the fourteen
studied metal-rich RR~Lyrae variables exhibit low
abundances of some chemical elements compared to
stars of other types with the same metallicity.

Unlike the other metal-rich RR~Lyrae stars,
KP~Cyg and TV~Lib have quite high abundances
of studied chemical elements in their atmospheres
(although the abundance of $\alpha$-elements in the former
star is low, like in thin disk stars). Marsakov et al.
(2018) drew attention to the overestimated [$\alpha$/Fe] ratio 
in TV~Lib, which is not typical for metal-rich
stars.

It would be interesting to conceive of the reason for
the special chemical composition of the atmospheres
of metal-rich RR~Lyrae variables of the galactic field:
both those, exhibiting anomalously low abundances
and those that differ from other metal-rich RR~Lyrae
stars in their high chemical element abundances.

Note that according to the probabilistic kinematic
criteria, only AA~Aql, BR~Aqr, TV~Lib and V~413~Oph
belong to the thick galactic disk, while all other
metal-rich RR~Lyrae stars from Table 2 belong to
the thin disk population.

\section {POSSIBLE EXPLANATIONS FOR THE
ORIGIN AND PROPERTIES OF
METAL-RICH RR~LYRAE STARS}

Therefore, in addition to the old metal-poor field
RR~Lyrae variables of the spherical subsystem (as
we previously thought of all RR~Lyrae stars), stars
with kinematic and orbital parameters typical of the
disk subsystems of the Galaxy and with low relative
abundances of some elements in their atmospheres
have been discovered. It is clear that the occurrence
of such unusual properties was provoked either by
the conditions of star birth, or by the external events
that the star experienced during its lifetime. Below,
referring to our own and other authors' studies of
recent years, we will present several possible explanations for 
the origin and specific properties of metal-rich 
RR~Lyrae variables. Note, however, that we
cannot exclude the possible inconsistency in determination 
of relative abundances of chemical elements in
RR~Lyrae stars and stars of other types.

\subsection {Increased Helium Abundance}

Marsakov et al. (2019) suggested that a possible reason 
for the appearance of such young, heavy-
element-rich RR~Lyrae stars could be high initial helium 
abundances in their precursors. Indeed, in this
case, the stars evolve faster and even with masses below 
solar, they manage to reach the stage of core helium 
burning during the existence of the  Galactic thin
disk, which is faster than over 10~billion years. The
populations of red giants and RR~Lyrae stars with
high helium abundances have already been discovered 
in the bulge, and some of them may well have
been moved to the solar vicinity by the disturbances 
caused by the inhomogeneities in the gravitational 
potential of the Galaxy. However, we cannot
explain the anomalies in the chemical composition of
such metal-rich RR~Lyrae stars by their radial migration 
from the Galactic center.

Our assumption that the low relative abundances
of $\alpha$-elements in them most likely indicate their
extragalactic origin (Marsakov et al., 2018, 2020)
seemed more logical, since by that time hardly any-
body doubted that giant spiral galaxies (like ours)
are formed as a result of merger of several less
massive ones at the early stages of the evolution of
the Universe (see the standard $\Lambda$CDM cosmological
model). In less massive satellite galaxies disrupted by
the Milky Way, the chemical evolution of interstellar
matter occurred differently. In addition, metal-rich
stars in such dwarf galaxies could have appeared later
than those in our Galaxy. We believe that the most
significant argument against the extragalactic origin
of such RR~Lyrae stars is the appearance in our Galaxy of 
stars (born in dwarf satellite galaxies) with 
circular (not elongated) orbits.

In addition, there is no evidence of the presence of
such RR~Lyrae stars in the stellar streams, the remnants 
of the already known destroyed dwarf galaxies.
Perhaps numerical modeling of the capture processes
can help to solve this problem.

\subsection {Binarity of Metal-Rich RR~Lyrae Variables}

Note one more possible reason for the appearance
of metal-rich RR~Lyrae stars. Bobrick et al.
(2024), using numerical simulations, tested the possibility 
of formation of metal-rich RR~Lyrae stars as a
result of evolution of close binary systems. According
to their hypothesis, at the red giant stage, a star fills
its Roche lobe and its outer envelope outflows onto
the companion star. As a result, the horizontal branch
stars, deprived of their outer envelope, become bluer
compared to similar single stars and, in this way, fall into
the instability strip. Bobrick et al. (2024) suggest
that all metal-rich RR~Lyrae stars have a long-period
companion.

Marsakov et al. (2019) discussed the probable
reasons for the small masses of metal-rich RR~Lyrae
stars ($0.51-0.60M_{\odot}$), if they are to be considered thin
disk objects. The following interpretation is possible.
In order to reach the horizontal branch and become an
RR~Lyrae variable during the lifetime of the thin disk
subsystem, the star has to lose a significant fraction
of its initial mass. The binarity of the RR~Lyrae stars
could explain the mass loss at the stage of mass
exchange between the components. But then all the
metal-rich RR~Lyrae stars of the thin disk should be
part of close binary systems, and we do not observe
this yet. Only two metal-rich RR~Lyrae stars are
known to be in binary systems so far.

Unfortunately, in such a scenario, it is difficult to
explain the existence of a relation between the binarity
and the anomalous chemical composition of stellar
atmospheres discussed in this paper.

\subsection {Incorrect Classification of Variables}

Some of the metal-rich RR~Lyrae stars may
have been incorrectly classified. In this regard,
two metal-rich stars are noteworthy, which
demonstrate significantly higher [Na/Fe], [Al/Fe],
[Sc/Fe], [Ti/Fe], [Ni/Fe], and [Y/Fe] ratios compared
to other metal-rich RR~Lyrae stars (see Table 2 and
the corresponding diagrams). These are the stars
KP~Cyg and TV~Lib.

KP~Cyg, with a rather high period for an RR~Lyrae
star of $0.^{d} 856$, can be classified as a type II Cepheid
(a BL~Hercules variable), rather than an RR~Lyrae
star (Andrievsky et al., 2010). Among the metal-rich
RR~Lyrae variables in our sample, KP~Cyg has the
highest metallicity ($\rm{[Fe/H]} = 0.15$) and is also highly
luminous compared to other metal-rich RR~Lyrae
stars (Marsakov et al., 2019).

TV~Lib is known as a very peculiar star, especially
due to its extremely short period ($0.^{d} 270$) (Molnar
et\,al.,~2016). The short period and the light curve
shape of TV~Lib may indicate a need for its different
classification, e.g. it may belong to $\delta$\,Scuti variables
(Kovacs and Karamiqucham,~2021). In addition, the
luminosity and period of TV~Lib suggest a younger
age than most of other field RR~Lyrae stars 
(Bono et\,al.,~1997).

Therefore, the properties of these two stars distinguish 
them from other metal-rich RR~Lyrae variables. 
If KP~Cyg and TVLib indeed belong to other
types of variables, then it makes no sense to consider
them when studying the properties of metal-rich
RR~Lyrae stars.

\subsection {RR~Lyrae Variables with Low Aluminum
Abundances}

Feuillet et\,al.~(2022) suggested that low relative 
aluminum abundances indicate an extragalactic 
origin for several metal-rich RR~Lyrae variables 
possessing the disk kinematics. This hypothesis 
was based on low aluminum abundances in the
stars of dwarf satellite galaxies. The authors considered 
stars to have been accreted if their aluminum
abundance is limited to a somewhat arbitrary value
of $\rm{[Al/Fe]} < -0.14$. The RR~Lyrae stars classified as
accreted in the cited paper are also investigated in
the present work. Feuillet et\,al.~(2022) were skeptical 
about the status of SW~And with an aluminum
abundance ($\rm{[Al/Fe]} = -0.11$), slightly exceeding the
limit they established. In our catalog (Marsakov et\,al.,~2018), 
the [Al/Fe] ratio for the SW~And variable, obtained 
by averaging over two measurements, turned
out to be $\rm{[Al/Fe]} = -0.07$, i.\,e. even higher. There
are no other metal-rich stars with known aluminum 
abundances that would satisfy the condition
$\rm{[Al/Fe]} < -0.14$ in the sample we are considering.

As a result, we have not added any new stars to the
list of accreted metal-rich aluminum-deficient
RR~Lyrae stars (Feuillet et\,al.,~2022), but we may
have removed the question of whether the SW~And
variable belongs to this category.

\section {CONCLUSION}

The paper considers the abundances of some
chemical elements in RR~Lyrae variable stars, as well
as their orbital parameters. Particular attention is
paid to the characteristics of RR~Lyrae stars with
$\rm{[Fe/H]} > -1$. In contrast to the earlier concept
of the RR~Lyrae stars as old, metal-poor stars of
the Galactic halo, it was found that these variables
belong to different galactic subsystems based on the
kinematic criteria. All the metal-rich RR~Lyrae stars
studied here demonstrate the orbital characteristics
typical mainly of objects in the thin disk and, much less
frequently, thick disk of the Galaxy.

An analysis of the relative abundances of several
chemical elements in the atmospheres of RR~Lyrae
variables was carried out. The abundances of aluminum, 
sodium, and nickel in most metal-rich variables 
proved to be lower than the average [Al/Fe],
[Na/Fe], and [Ni/Fe] values in the field stars of the
thin disk. Since the comparison was made with both
stationary and variable field stars, it is unlikely that
the low [Al/Fe], [Na/Fe], and [Ni/Fe] are associated
with non-stationary processes occurring in the atmospheres 
of these stars. The result on the anomalously
low abundances of aluminum, sodium, and nickel in
the metal-rich RR~Lyrae stars, supplemented by
the conclusions of our previous studies (Marsakov et\,al.,~2018; 
Gozha et\,al.,~2020) on the low [Mg/Fe],
[Si/Fe], [Ca/Fe], [Sc/Fe], [Ti/Fe], and [Y/Fe] ratio
values in metal-rich RR~Lyrae variables compared to
other types of stars of similar metallicity, points at
the unusual nature of these objects. It turned out
that twelve out of the fourteen metal-rich RR~Lyrae
variables studied in the paper reveal the [X/Fe] ratios
below the expected values for at least a few chemical
elements. And in the atmospheres of the metal-rich
RR~Lyrae stars DM~Cyg, CN~Lyr, and V~445~Oph, we
have discovered low abundances of all the chemical
elements studied in them relative to the comparison
stars of the same metallicity.

It can be assumed that the observed abundances
of the elements under consideration in metal-rich
RR~Lyrae stars may be a consequence of their formation 
from matter, the chemical evolution history
of which differs from the evolution history of the host
interstellar matter of most comparison stars. These
RR~Lyrae stars may have been captured from the
satellite galaxies of the Milky Way, which is confirmed
by the identity of the chemical composition 
of metal-rich RR~Lyrae stars and stars
of some dwarf satellite galaxies (see Hasselquist et\,al.,~2021). 
If we assume the extragalactic origin of some
metal-rich RR~Lyrae stars, then we can talk about
the presence of this type of stars not only among the
objects, genetically related to a single protogalactic
cloud, but also about their presence in the subsystem
of the accreted halo.

We could try to explain the unusual chemical 
composition and low masses of metal-rich RR~Lyrae stars
by the binarity of such stars. However, there are
currently too few known RR~Lyrae variables with high
metallicity that are part of binary systems. Therefore,
such an assumption cannot be considered valid and it
requires further verification.

An excessive helium abundance in the progenitors
of metal-rich RR~Lyrae stars would lead to a faster
evolution of such stars and their entry into the instability 
strip. Although there are observations 
indicating a certain number of such stars in the center
of the Galaxy, the question of the deficit of individual
chemical elements in their atmospheres still remains
open.

Two stars from our list of RR~Lyrae variables
with $\rm{[Fe/H]} > -1$, KP~Cyg and TV~Lib demonstrate
chemical properties that distinguish them from other
metal-rich RR~Lyrae stars. One can suspect an
incorrect classification of these stars by the variability
type. Then the conclusions about the properties of
the metal-rich RR~Lyrae stars may be distorted
by the presence of different type of variables in the
studied sample.

The kinematic and orbital parameters indicating
their belonging to the disk subsystems and the
anomalous chemical composition of the metal-rich
RR~Lyrae variables indicate the unusual nature of
these stars and encourage us to continue our research.

\section*{ACKNOWLEDGMENTS}
The authors are grateful to the reviewer for useful
comments that contributed to the improvement of the
paper. The work used data from the Gaia mission
of the European Space Agency 
(ESA, https://www.cosmos.esa.int/gaia), prepared by the Gaia
Data Processing and Analysis Consortium (DPAC,
https://www.cosmos.esa.int/web/gaia/dpac/consortium). 
The study also made use of the SIMBAD database 
(https://simbad.cds.unistra.fr/simbad).

\section*{FUNDING}

The study was carried out at the Southern Federal
University with the financial support of the Ministry of
Science and Higher Education of the Russian Federation 
(government contract GZ0110/23-10-IF).

\section*{CONFLICT OF INTEREST}

The authors declare no conflict of interest.

\renewcommand{\refname}{REFERENCES}

\newpage

\begin{figure*}
\centering
\includegraphics[angle=0,width=0.99\textwidth,clip]{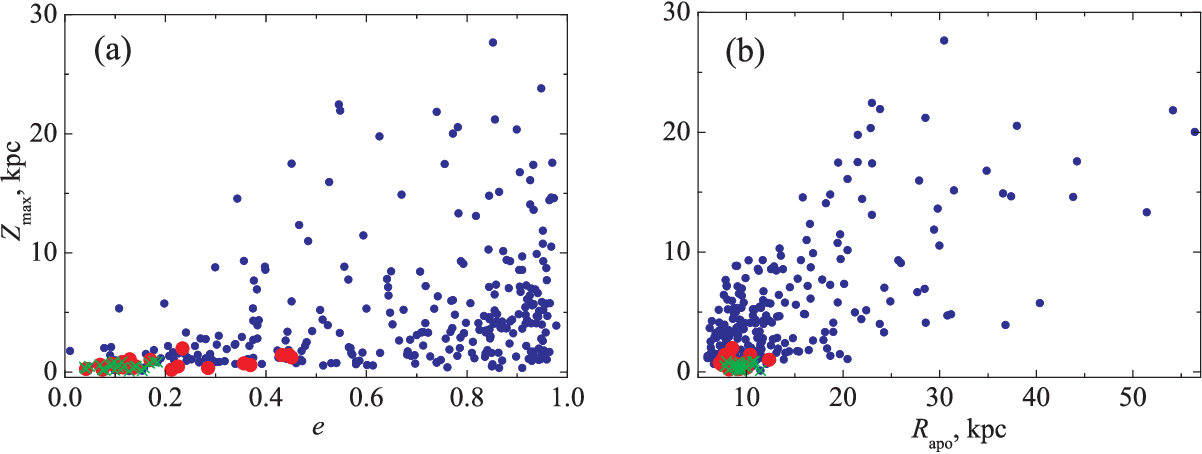}
\caption{RR~Lyrae variables from the catalog (Prudil et\,al.,~2020) 
         in the diagrams ``$e$ -- $Z_{max}$'' (a) and 
         ``$R_{apo}$ -- $Z_{max}$'' (b). The blue circles denote 
         all the stars from the catalog of Prudil et\,al.~(2020), 
         red---RR~Lyrae stars with $\rm{[Fe/H]} > -1$, selected 
         from the catalog of Marsakov et\,al.~(2018), while the 
         green crosses are the RR~Lyrae stars of the disk 
         according to Prudil et\,al.~(2020).}
\label{fig1}
\end{figure*}

\newpage

\begin{figure*}
\centering
\includegraphics[angle=0,width=0.99\textwidth,clip]{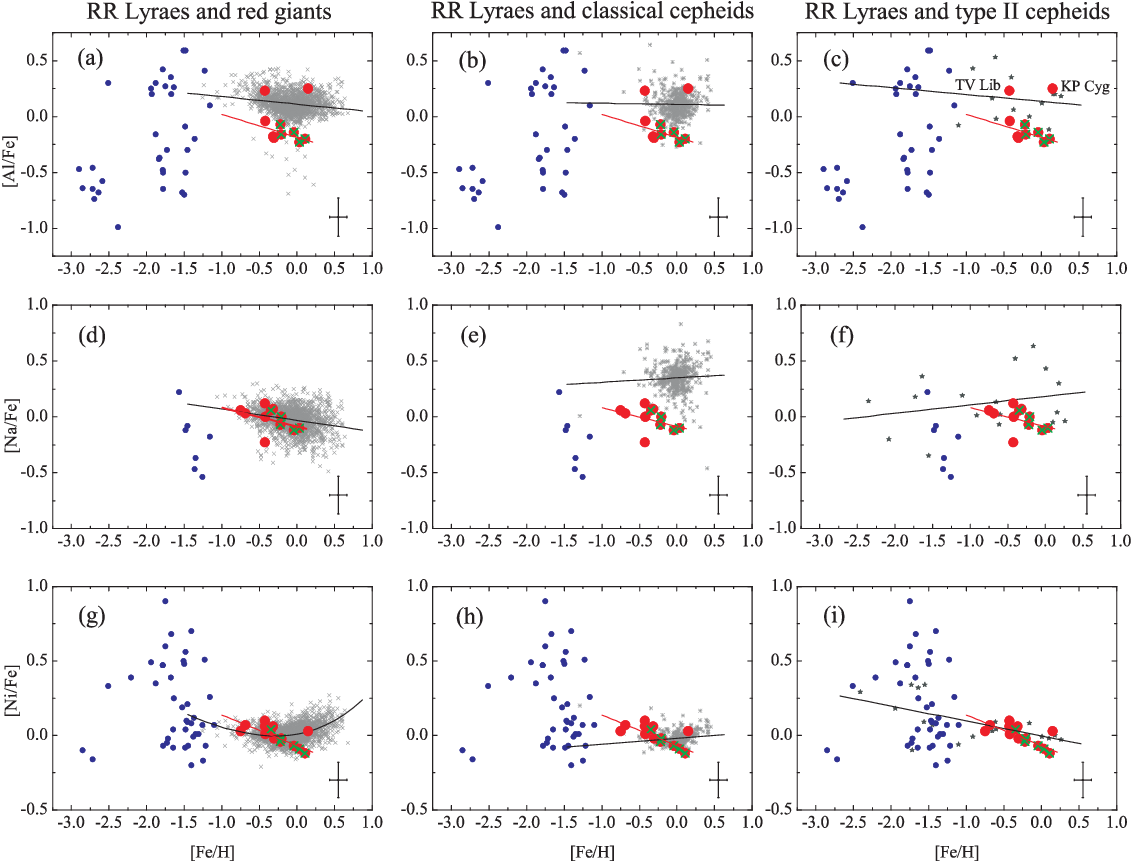}
\caption{Relative abundances of aluminum, sodium, and nickel as a 
         function of metallicity for the RR~Lyrae variables from 
         the Galactic field and the comparison stars. The RR~Lyrae 
         stars from the catalog by Marsakov et\,al.~(2018) with 
         $\rm{[Fe/H]} < -1$ are denoted by the blue circles, and 
         those with $\rm{[Fe/H]} > -1$ --- by the red circles. The 
         green crosses denote the disk RR~Lyrae stars from 
         Prudil et\,al.~(2020). The comparison stars: gray skew 
         crosses (a), (d), (g) show 1918 red giants from 
         Hawkins et\,al.~(2016); gray snowflakes (b), (e), (h) 
         are 435~Cepheids from Luck~(2018); gray stars (c), (f), 
         (i)--23 type II Cepheids (W~Virgo and BL~Hercules-type) 
         from Kovtyukh et al. (2018). The red lines describe the 
         trends in the relative abundance variation of a chemical 
         element with increasing metallicity for the RR~Lyrae 
         stars with $\rm{[Fe/H]} > -1$, the black lines-for the 
         comparison stars. The [Fe/H] and [X/Fe] error bars 
         for the RR~Lyrae stars are shown.}
\label{fig2}
\end{figure*}

\newpage

\begin{figure*}
\centering
\includegraphics[angle=0,width=0.99\textwidth,clip]{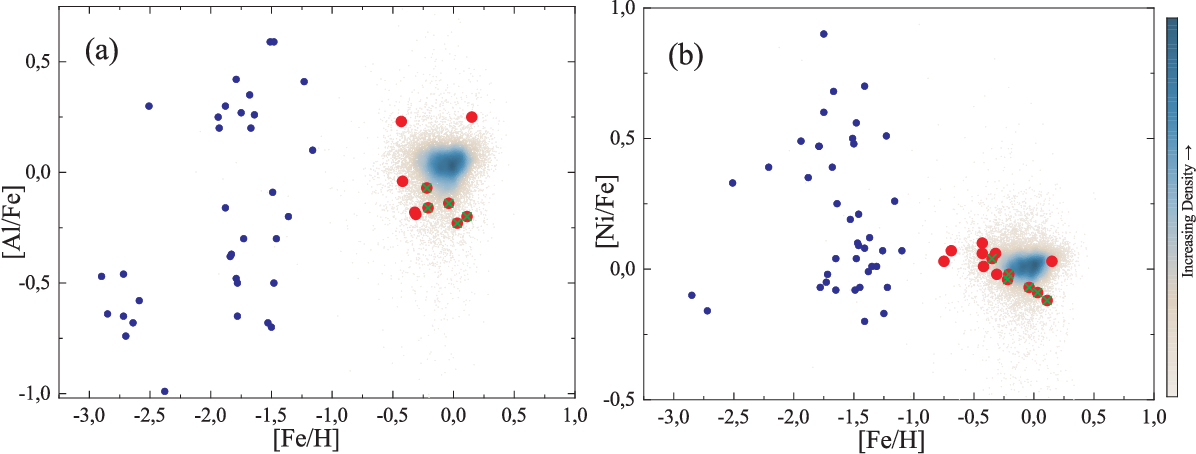}
\caption{The ``[Fe/H] -- [Al/Fe]'' (a) and ``[Fe/H] -- [Ni/Fe]''
         (b) diagrams for the RR~Lyrae stars from the catalog 
         by Marsakov et\,al.~(2018) and the stars from the 
         catalog by J\"onsson et\,al.~(2020). The markings 
         for the RR~Lyrae stars are the same as in Fig~1. 
         The stars from the J\"onsson et\,al. catalog~(2020) 
         are indicated by dots (the density of dots on the 
         diagram is shown in color).}
\label{fig3}
\end{figure*}

\newpage
\begin{landscape}
\renewcommand{\baselinestretch}{0.85}
\begin{table}
\caption {Metallicity and relative nickel abundances in RR~Lyrae 
variables of the Galactic field \footnotesize {({1 --~Andrievsky et\,al.~(2010),
2 --~Clementini et\,al.~(1995), 3 --~For and Sneden~(2010), 
4 --~For et\,al.~(2011), 5 --~Govea et\,al.~(2014),, 6 --~Hansen et\,al.~(2011),
7 --~Kolenberg et\,al.~(2010), 8 --~Liu et\,al.~(2013), 
9 --~Pancino et\,al. (2015)}})} 
\medskip
\begin{tabular}{c|c|c|c||c|c|c|c}
\hline
Star             &[Fe/H],&[Ni/Fe],&References&   Star       &[Fe/H],&[Ni/Fe],&References\\
                 &  dex  &  dex   & [Ni/Fe]  &              &  dex  &  dex&[Ni/Fe]\\
\hline                
SW And           & -0.22 & -0.04  &  [2, 8]  &DH Hya        & -1.53 &  0.19&[9]\\
CI And           & -0.43 &  0.06  &   [8]    &DT Hya        & -1.23 &  0.51&[4]\\
DR And           & -1.37 &  0.12  &   [9]    &V Ind         & -1.45 & -0.07&[9]\\
WY Ant           & -1.88 &  0.35  &   [4]    &SS Leo        & -1.75 &  0.90&[9]\\
XZ Aps           & -1.79 &  0.47  &   [4]    &TV Lib        & -0.43 &  0.10&[8]\\
BS Aps           & -1.48 &  0.56  &   [4]    &RR Lyr        & -1.49 & -0.08&[2,7,8]\\
AA Aql           & -0.32 &  0.06  &   [8]    &CN Lyr        & -0.04 & -0.07&[8]\\
BR Aqr           & -0.69 &  0.07  &   [8]    &IO Lyr        & -1.35 &  0.01&[8]\\
X Ari            & -2.51 &  0.33  &   [9]    &KX Lyr        & -0.42 &  0.01&[8]\\
RS Boo           & -0.21 & -0.02  &   [8]    &Z Mic         & -1.51 &  0.50&[4]\\
ST Boo           & -1.73 & -0.05  &   [2]    &RV Oct        & -1.64 &  0.25&[4]\\
TW Boo           & -1.47 &  0.10  &   [9]    &UV Oct        & -1.75 &  0.60&[4]\\
BPS CS 22881-039 & -2.72 & -0.16  &   [6]    &V 413 Oph     & -0.75 &  0.03&[8]\\
BPS CS 22940-070 & -1.41 &  0.70  &   [3]    &V 445 Oph     &  0.11 & -0.12&[2,8]\\
BPS CS 30317-056 & -2.85 & -0.10  &   [6]    &AO Peg        & -1.26 &  0.07&[8]\\
UZ CVn           & -2.21 &  0.39  &   [9]    &DH Peg        & -1.31 &  0.01&[8]\\
YZ Cap           & -1.50 &  0.48  &   [5]    &VW Scl        & -1.22 & -0.07&[9]\\
RR Cet           & -1.48 &  0.04  &  [2, 8]  &VY Ser        & -1.78 & -0.07&[2,8]\\
RX Cet           & -1.38 & -0.01  &   [9]    &V 440 Sgr     & -1.16 &  0.26&[2]\\
U Com            & -1.41 &  0.08  &   [9]    &V 1645 Sgr    & -1.94 &  0.49&[4]\\
DM Cyg           &  0.03 & -0.09  &   [8]    &BK Tuc        & -1.65 &  0.04&[9]\\
KP Cyg           &  0.15 &  0.03  &   [1]    &TYC 4887-622-1& -1.79 &  0.47&[5]\\
DX Del           & -0.31 & -0.02  &   [8]    &RV UMa        & -1.25 & -0.17&[9]\\
AE Dra           & -1.46 &  0.21  &   [9]    &TU UMa        & -1.41 & -0.20&[9]\\
BK Eri           & -1.72 & -0.02  &   [9]    &CD Vel        & -1.67 &  0.68&[4]\\
SZ Gem           & -1.65 & -0.08  &   [9]    &UV Vir        & -1.10 &  0.07&[9]\\
TW Her           & -0.35 &  0.04  &   [8]    &AS Vir        & -1.68 &  0.39&[4]\\
VX Her           & -1.46 &  0.09  & [2,8,9]  &              &       &   &\\
\hline 
\end{tabular}\\
\end{table}

\newpage

\begin{table}
\caption {Relative abundances of some elements in metal-rich 
($\rm{[Fe/H]} > -1$) RR~Lyrae variables}
\medskip
\begin{tabular}{c|c|c|c|c|c|c|c}
\hline
Star      & [Fe/H] & [Na/Fe] & [Al/Fe] & [Sc/Fe] & [Ti/Fe] & [Ni/Fe] & [Y/Fe]\\
\hline
\hline
SW And    &  -0.22 &  -0.07  &  -0.07  & -0.40  &   -0.16  &  -0.04  & -0.53\\
\hline
CI And    &  -0.43 &  -0.23  &   --    & -0.28  &   -0.05  &   0.06  &  --\\
\hline
AA Aql    &  -0.32 &   0.07  &  -0.18  & -0.22  &    0.06  &   0.06  & -0.39\\
\hline
BR Aqr    &  -0.69 &   0.03  &   --    & -0.45  &    0.14  &   0.07  & -0.73\\
\hline
RS Boo    &  -0.21 &   0.00  &  -0.16  & -0.37  &   -0.12  &  -0.02  & -0.50\\
\hline
DM Cyg    &   0.03 &  -0.10  &  -0.23  & -0.48  &   -0.24  &  -0.09  & -0.53\\
\hline
KP Cyg    &   0.15 &    --   &   0.25  & -0.02  &   -0.08  &   0.03  & -0.02\\
\hline
DX Del    &  -0.31 &    --   &  -0.19  & -0.35  &   -0.13  &  -0.02  & -0.37\\
\hline
TW Her    &  -0.35 &   0.06  &   --    & -0.35  &   -0.17  &   0.04  & -0.73\\
\hline 
TV Lib    &  -0.43 &   0.12  &   0.23  &  0.12  &    0.24  &   0.10  & -0.13\\
\hline
CN Lyr    &  -0.04 &  -0.12  &  -0.14  & -0.32  &   -0.14  &  -0.07  & -0.36\\
\hline
KX Lyr    &  -0.42 &   0.00  &  -0.04  & -0.36  &   -0.06  &   0.01  & -0.47\\
\hline
V 413 Oph &  -0.75 &   0.06  &   --    & -0.27  &    0.15  &   0.03  & -0.49\\
\hline 
V 445 Oph &  0.11  &    --   &  -0.20  & -0.46  &   -0.30  &  -0.12  & -0.42\\
\hline
\end{tabular}\\
\end{table}


\end{landscape}

\end{document}